\documentclass[10pt]{article} 
\usepackage{amssymb}
\usepackage{amsmath}
\usepackage{amsthm}
\usepackage{latexsym} 
\usepackage[dvips]{epsfig}
\usepackage{mathrsfs}
\usepackage{eufrak}

\theoremstyle{plain}
 
\newtheorem{lemma}{Lemma}
\newtheorem{theorem}{Theorem}
\newtheorem{assumption}{Assumption}

\newtheorem*{definition}{Definition}

\def\O{\mathcal{O}}

\def\d{\mbox{d}}

\font\SYM=msbm10 
\newcommand{\Real}{\mbox{\SYM R}}
\newcommand{\Complex}{\mbox{\SYM C}}

\font\tenscr=rsfs10 scaled1100
\font\sevenscr=rsfs7 
\font\fivescr=rsfs5 
\skewchar\tenscr='177 
\skewchar\sevenscr='177
\skewchar\fivescr='177 
\newfam\scrfam 
\textfont\scrfam=\tenscr
\scriptfont\scrfam=\sevenscr 
\scriptscriptfont\scrfam=\fivescr

\newcommand{\TT}[3]{T_{#1 \phantom{#2} #3}^{\phantom{#1} #2}}

\begin{document}


\title{\textbf{Time asymmetric spacetimes near null and spatial infinity. II. Expansions of developments of initial data sets with non-smooth conformal metrics.}}

\author{Juan Antonio Valiente Kroon \thanks{E-mail address:
 {\tt jav@ap.univie.ac.at}} \\
 Institut f\"ur Theoretische Physik,\\ Universit\"at Wien,\\
Boltzmanngasse 5, A-1090 Wien,\\ Austria.}

\maketitle

\begin{abstract}
This article uses the conformal Einstein equations and the conformal
representation of spatial infinity introduced by Friedrich to analyse
the behaviour of the gravitational field near null and spatial
infinity for the development of initial data which are, in principle,
non-conformally flat and time asymmetric. This article is the
continuation of the investigation started in Class. Quantum Grav. 21
(2004) 5457-5492, where only conformally flat initial data sets were
considered. For the purposes of this investigation, the conformal
metric of the initial hypersurface is assumed to have a very
particular type of non-smoothness at infinity in order to allow for
the presence of non-Schwarzschildean initial data sets in the class
under study. The calculation of asymptotic expansions of the
development of these initial data sets reveals ---as in the
conformally flat case--- the existence of a hierarchy of obstructions
to the smoothness of null infinity which are expressible in terms of
the initial data. This allows for the possibility of having spacetimes
where future and past null infinity have different degrees of
smoothness. A conjecture regarding the general structure of the
hierarchy of obstructions is presented.

\end{abstract}

\textbf{Pacs: 04.20.Ha, 04.20.Ex}

\section{Introduction}

This article is the second part of an investigation of the
gravitational field near null and spatial infinity arising as the time
development of initial data sets which are not time symmetric ---that
is, the second fundamental form of the initial data set is
non-vanishing. Part I of this study was concerned with the analysis of
initial data sets for which the 3-metric of the initial hypersurface
---its first fundamental form--- was assumed to be conformally flat
\cite{Val04e}. The main result of part I was the uncovering of a time
asymmetric hierarchy of obstructions to the smoothness of null
infinity, which if vanishing, would imply that the initial data set
under consideration is asymptotically Schwarzschildean to a certain
order. The obstructions found in part I are the direct generalisations
of those found in \cite{Val04a} for time symmetric, conformally flat
initial data. In hindsight, the analysis performed in \cite{Val04a}
could be readily generalised in two directions: (i) removing the
conformal flatness assumption, but keeping the time symmetry; (ii)
keeping the conformal flatness of the initial 3-metric, but taking away
the time symmetry; (iii) doing without the conformal flatness and time
symmetry. The generalisation (i) was carried out in \cite{Val04d},
where a hierarchy of obstructions implying asymptotic staticity of the
initial data near infinity was found. As mentioned earlier,
generalisation (ii) was carried out in part I.

This article is essentially concerned with generalisation (iii). That
is, we want to deduce asymptotic expansions of the gravitational field
near null and spatial infinity for the developments of time asymmetric
initial data sets which are not conformally flat. From the results in
\cite{Val04a,Val04d} and part I, we expect that the stationary
spacetimes will play a crucial role in our analysis. As we are
constructing our asymptotic expansions out of Cauchy initial data
sets, we are in the need of considering a class of initial data big
enough to contain strictly stationary data ---i.e. not only disguised
static data. From the results concerning the non-existence of
conformally flat initial slices in the Kerr and other stationary
solutions, it is clear that with conformally flat data, one would only
be dealing with ---more or less sophisticated--- Schwarzschildean data
\cite{Val04c}. Thus, one has to move away from conformal flatness. But
this is not enough, for in \cite{Dai01c} it was shown that the
conformal metric of Kerrian initial data is actually non-smooth ---a
feature that is bound to pervade all stationary solutions with
non-vanishing angular momentum.

The general strategy of this article can be summarised as follows:
given a sufficiently wide class of initial data sets, we calculate
asymptotic expansions for the initial data of the conformal Einstein
equations. These expansions are, in turn, fed into the conformal
evolution equations and used to obtain in a recursive way asymptotic
expansions of the development of the initial data sets. These
expansions describe the behaviour of the resulting spacetime in the
region close to null and spatial infinity. The recursive nature of the
calculations is quite amenable for the implementation on a computer
algebra (CA) system. Indeed, the crucial results given in this article
have been calculated using scripts written in the system {\tt Maple V}.

As in part I, the asymptotic expansions show generically a certain
type of logarithmic non-smoothness at the sets where null infinity and
spatial infinity ``touch''. Because of the hyperbolic nature of the
evolution part of the conformal field equations it is quite likely
that this non-smoothness will be propagated along the generators of
null infinity. Hence, the conformal completion will be non-smooth. A
remarkable feature of our set up is that it
allows us to identify the parts of the initial data responsible for
these logarithmic divergences. As a consequence, one obtains a
hierarchy of quantities expressible in terms of the initial data
---henceforth to be referred to as an \emph{obstruction}--- whose
vanishing would eliminate a certain type of logarithmic
divergences.The hierarchy exhibits a subtle structure. Actually,
the regularity observed in our calculations permits to infer what the
general behaviour of the obstructions to any order should be.

In part I, the relative simplicity of the conformal data permitted to
show that the vanishing of the obstructions to a certain order leads
to an initial data set which is asymptotically Schwarzschildean ---but
not necessarily time symmetric. In the case of the current article,
the increased complexity of the initial data sets considered,  and of the
resulting expansions preclude us from deducing an analogous condition
with regards to stationarity. That is, if one were able to carry the
expansions to high enough order one would expect to be able to deduce
asymptotic stationarity from the hierarchy of obstructions. An
analysis of how stationary data and their asymptotic expansions fit
in the framework of the cylinder at spatial infinity will be pursued
elsewhere.

As a continuation to \cite{Val04e}, this article is
completely consistent with the notation, conventions and
nomenclature of part I. The main features of our set up, including the
construction of the cylinder at infinity and the use of 2-spinors have
been discussed there and the reader will be duly referred to it
whenever it is necessary. The article is structured as follows: in
section 2, we discuss the existence of solutions to the constraint
equations which are expandable in the asymptotic region in terms of
powers of $1/r$. The discussion of this section is based in partially
unpublished results due to S.Dain. As we are using systematically a
2-spinor formalism, in section 3, we discuss the effects of our class
of initial data on the solutions of the structure equations. In
section 4, it is shown how to fix the conformal gauge in the initial
data. This is done by resorting to a certain gauge based on conformal
geodesics on the initial hypersurface (the cn-gauge). This procedure
of fixing the conformal gauge is important if one wants to
identify the parts of the freely specifiable data which are pure
gauge. Section 5 is concerned with the expansions of solutions of the
non-conformally flat momentum constraint. Section 6 does likewise but
now with the solutions of the Hamiltonian constraint. Section 7
discusses how a certain regularity condition first found by H. Friedrich
can be implemented in our set up. Section 8 is concerned with the
solutions of the transport equations implied by the conformal Einstein
equations on the cylinder at spatial infinity from which our
asymptotic expansions are deduced. Section 9 speculates about the
general behaviour of the hierarchy of obstructions and the associated
logarithmic divergences of the solutions to the transport equations at
the sets where null infinity ``touches'' spatial infinity. Finally,
section 10 provides some conclusions.

In order to carry out the calculations here described, a number of
assumptions regarding the initial data set have been made. These are
clearly numbered from 1 to 5.

\subsection{General conventions}
Throughout, we shall fully follow the conventions used in part I. In
particular, let $(\widetilde{\mathcal{M}},\widetilde{g}_{\mu\nu})$
denote a spacetime arising as the development of some Cauchy initial
data
$(\widetilde{\mathcal{S}},\widetilde{h}_{\alpha\beta},\widetilde{\chi}_{\alpha\beta})$.
Tilded quantities will refer to quantities in the \emph{physical}
spacetime, whereas untilded ones will denote generically quantities on
an \emph{unphysical} ---i.e. conformally rescaled--- spacetime.  The
indices $\mu$, $\nu$, $\lambda,\ldots$ (second half of the Greek
alphabet) are spacetime indices taking the values $0,\ldots,3$; while
$\alpha$, $\beta$, $\gamma,\dots$ are spatial ones with range
$1,\ldots,3$. The latin indices $a$, $b$, $c,\ldots$ will be used in
spatial expressions which are valid for a particular coordinate system
(usually a Cartesian normal one) and take the values $1,\ldots,3$. The
indices, $i$, $j$, $k,\ldots$ are spatial frame indices ranging
$1,\ldots,3$, while $A$, $B$, $C,\ldots$ will be spinorial indices
taking the values $0$, $1$. Because of the use of spinors, the
signature of $\widetilde{g}_{\mu\nu}$ will be taken to be $(+,-,-,-)$,
and the 3-dimensional metric $\widetilde{h}_{\alpha\beta}$ will be
negative definite.

We shall be concentrating in a particular asymptotically flat end of
the manifold $\mathcal{S}$. Without loss of generality, we shall assume
that there is only one asymptotic end. Let $\{y^a\}$ denote
coordinates adapted to that end in the sense that
$\widetilde{h}_{ab}=-\delta_{ab}+\O(1/|y|)$ as $|y|\rightarrow
\infty$. We shall also assume
$(\widetilde{\mathcal{S}},\widetilde{h}_{\alpha\beta})$ to be
\emph{asymptotically Euclidean and regular} in the sense described in
part I. Accordingly, let $(\mathcal{S},h_{\alpha\beta})$ denote the
3-dimensional, orientable, smooth compact Riemannian manifold and
$i\in \mathcal{S}$ the point whose suitable punctured neighbourhoods
correspond to the asymptotically flat end of
$(\widetilde{\mathcal{S}},\widetilde{h}_{\alpha\beta})$. The manifold
$\widetilde{\mathcal{S}}$ is identified in a standard way with
$\mathcal{S}\setminus\{i\}$. Most of our discussion shall be carried
out in a sufficiently small neighbourhood $\mathcal{B}_a(i)$ of radius
$a$ centred on $i$. The neighbourhood $\mathcal{B}_a(i)$ will be
assumed to be geodesically convex.  Unless otherwise stated, $\{x^a\}$
denote some normal coordinates with origin at $i$ based on a
$h$-orthonormal frame $\{e_j\}$.

A crucial feature of our analysis is that the region of spacetime near
null and spatial infinity can be described by means of a certain
manifold $\mathcal{M}_{a,\kappa}$ given by
\begin{equation}
\mathcal{M}_{a,\kappa}=\bigg\{(\tau,q) \bigg| q\in \mathcal{C}_{a,\rho},-\frac{\omega(q)}{\kappa(q)}\leq \tau \leq \frac{\omega(q)}{\kappa(q)}\bigg\}, 
\end{equation}
where $\mathcal{C}_{a,\kappa}=\rho^{1/2}\mathcal{C}_a$ and
$\mathcal{C}_a$ is the blow up of the neighbourhood
$\mathcal{B}_a(i)$. The conformal factor associated with this
representation of the ``unphysical'' spacetime is given by
\begin{equation}
\Theta=\kappa^{-1}\Omega \left(1-\tau^2\frac{\kappa^2}{\omega^2}\right),
\end{equation}
where $\Omega$ is the conformal factor of the initial hypersurface and
\begin{equation}
\omega=\frac{2\Omega}{\sqrt{|D_\alpha \Omega D^\alpha\Omega|}}.
\end{equation}
 The function $\kappa$ containing the remaining freedom in our setting
 will be set equal to
\begin{equation}
\kappa=\rho,  
\end{equation}
in order to ease our calculations. Accordingly, the critical sets, 
$\mathcal{I}^\pm$, where null infinity ``touches'' spatial infinity are
given by the point where $\rho=0$ and $\tau=\pm 1$.

\bigskip
In order to ease the presentation and understanding of the results
described in this article, we shall be making the following assumption

\setcounter{assumption}{-1}

\begin{assumption}[axial symmetry]
The initial data sets
$(\widetilde{\mathcal{S}},\widetilde{h}_{\alpha\beta},\widetilde{\chi}_{\alpha\beta})$
to be considered will be assumed to be axially symmetric.
\end{assumption}

It follows from well known results ---see e.g. \cite{Mon75}--- that the
development of the initial data will inherit the axial symmetry. It
must be stressed that there is no fundamental restriction implied by
the latter assumption. All the results presented in this article can
be extended naturally ---aside from the associated increase of
computational complexity--- to the non-axially symmetric case.

\section{Existence of solutions to the constraint equations with non-smooth conformal metrics}
\label{section:existence}

As discussed in part I, for technical reasons, we are interested in 
solutions
$(\widetilde{h}_{\alpha\beta},\widetilde{\chi}_{\alpha\beta})$ to the
Einstein constraint equations
\begin{subequations}
\begin{eqnarray}
&& \widetilde{r}-\widetilde{\chi}^2+\widetilde{\chi}_{\alpha\beta}\widetilde{\chi}^{\alpha\beta}=0, \label{constraint_1} \\
&& \widetilde{D}^\alpha\widetilde{\chi}_{\alpha\beta}-\widetilde{D}_\beta\widetilde{\chi}=0, \label{constraint_2}
\end{eqnarray}
\end{subequations}
which on an asymptotic end of the initial hypersurface
$\widetilde{\mathcal{S}}$ admit expansions of the form
\begin{equation}
\label{asymptotic}
\widetilde{h}_{\alpha\beta}\sim\left(1+\frac{2m}{|y|}\right)\delta_{\alpha\beta}+\sum_{k\geq2}\frac{\widetilde{h}^{(k)}_{\alpha\beta}}{|y|^k}, \quad \widetilde{\chi}_{\alpha\beta}\sim \sum_{k\geq 2}\frac{\widetilde{\chi}^{(k)}_{\alpha\beta}}{|y|^k},
\end{equation}
where $\widetilde{h}^{(k)}_{\alpha\beta}$ and $\widetilde{\chi}^{(k)}_{\alpha\beta}$ are smooth functions on the sphere.

\bigskip
We shall perform our analysis not in the \emph{physical
spacetime} which arises as the time development of
$(\widetilde{\mathcal{S}},\widetilde{h}_{\alpha\beta},\widetilde{\chi}_{\alpha\beta})$,
but in a conformally rescaled version thereof ---the \emph{unphysical
spacetime}. Writing as usual
\begin{equation}
g_{\mu\nu}=\Omega^2\widetilde{g}_{\mu\nu},
\end{equation} 
we make the following assumption (which was also made in part I):

\begin{assumption}[maximal initial data]
It shall be assumed that
\begin{equation}
\Sigma=0, \quad \Omega\chi=\widetilde{\chi}=0 \mbox{ on }
\widetilde{\mathcal{S}},
\end{equation}
 where $\Sigma$ denotes the derivative of
$\Omega$ along the future $g-$unit normal of
$\widetilde{\mathcal{S}}$.
\end{assumption}

Under the latter assumption, the constraint equations
(\ref{constraint_1})-(\ref{constraint_2}) imply on the unphysical
manifold $\mathcal{S}$ the equations
\begin{subequations}
\begin{eqnarray}
&& \left(D_\alpha D^\alpha -\frac{1}{8}r\right)\vartheta =\frac{1}{8}\psi_{\alpha\beta}\psi^{\alpha\beta}\vartheta \mbox{ with } \vartheta=\Omega^{-1/2}, \label{Hamiltonian}\\
&& D^\alpha\left(\psi_{\alpha\beta}\right)=0, \label{momentum}
\end{eqnarray}
\end{subequations}
where $D$ is the connection associated with the conformal 3-metric
\begin{equation}
\label{h_alphabeta}
h_{\alpha\beta}=\Omega^2\widetilde{h}_{\alpha\beta},
\end{equation}
and the 3-tensor $\psi_{\alpha\beta}$ is given by
\begin{equation}
\label{psi_alphabeta}
\psi_{\alpha\beta}=\Omega^{-1}\widetilde{\chi}_{\alpha\beta}=\vartheta^2\widetilde{\chi}_{\alpha\beta}.
\end{equation} 
The equations (\ref{Hamiltonian}) and (\ref{momentum}) are to be
supplemented with the asymptotic (boundary) conditions
\begin{equation}
\label{boundary}
|x|\vartheta\rightarrow 1, \quad \psi_{\alpha\beta}=\O\left(\frac{1}{|x|^4}\right), \quad \mbox{ as } |x|\rightarrow 0,
\end{equation}
where $\{x^a\}$ are normal coordinates with origin at $i$.

Conditions for the existence of solutions to the equations
(\ref{Hamiltonian})-(\ref{momentum}) with the above boundary
conditions, and such that the corresponding physical fields have the
asymptotic behaviour (\ref{asymptotic}) have been given in
\cite{DaiFri01} under the assumption that the conformal metric
$\widetilde{h}_{\alpha\beta}$ is smooth in a neighbourhood
$\mathcal{B}_a(i)$ of infinity with respect to some normal
coordinates. Unfortunately, it seems that the assumption of smoothness
of the conformal metric is not enough to discuss stationary
spacetimes. In \cite{Dai01c} it has been shown that for Kerrian data
derived from the Kerr spacetime in Boyer-Lindquist coordinates,
\begin{equation}
\label{non_smooth}
h_{ab}=h^{(I)}_{ab}+|x|^3h^{(II)}_{ab},
\end{equation}
where $h^{(I)}_{ab}$ and $h^{(II)}_{ab}$ are analytic in
$\mathcal{B}_a(i)$. The important fact to be noted here is that
$|x|^3\in C^{2,\alpha}(\mathcal{B}_a(i))$, whence also $h_{ab}\in
C^{2,\alpha}(\mathcal{B}_a(i))$. In the light of the results given in
\cite{Dai01b,Val04c} it is very unlikely that there are slices in the
Kerr solution with a better smoothness at infinity. Now, stationary
spacetimes are, to leading order, Kerrian \cite{BeiSim80}, and thus,
similar non-smoothness is to be expected from any other stationary
solution with non-vanishing angular momentum.

Existence results for solutions to the equations (\ref{Hamiltonian})
and (\ref{momentum}) under assumption of a conformal metric of the
form given in equation (\ref{non_smooth}) and which are expandable in
powers of $|x|$ are only available under the premise of axial symmetry
\cite{Dai04a,Dai04b}. Nevertheless, results for the non-axisymmetric
case seem plausible.

Dain's existence results use the fact that for axially symmetric
initial data sets it is possible to explicitly write solutions to the
momentum constraint. These solutions are calculated from freely a 
specifiable function which acts as a potential. So, if the metric
$h_{ab}$ has an axial Killing vector $\eta^a$, then a solution of the
momentum constraint, equation (\ref{momentum}) is given by
\begin{equation}
\psi^{ab}=\frac{2\psi^{(a}\eta^{b)}}{h_{cd}\eta^c\eta^d}, \label{axial_psi_ab}
\end{equation}
 where 
\begin{equation}
\psi^a=\frac{1}{h_{de}\eta^d\eta^e}\epsilon^{abc}\eta_b D_c\omega, \quad \mathcal{L}_\eta\omega=0. \label{psi_a} 
\end{equation}
The potential $\omega$ in the last equation is calculated from a spin
0 real function $\lambda$ and a constant $J^z$ via
\begin{equation}
\omega=\eth^2\lambda\sin^2\theta + J^z(-3\cos\theta+\cos^3\theta), \label{omega}
\end{equation}
where $\eth$ is the standard NP ``eth'' operator.

It turns out that solutions to the momentum constraint constructed by
the aforediscussed procedure satisfy the regularity condition given
by Dain \& Friedrich in \cite{DaiFri01}. Indeed,

\begin{theorem}[Dain, 2003]
\label{existence_psi}
Assume that a metric of the form (\ref{non_smooth}) has an axial
Killing vector $\eta^a$. Let $\psi^{ab}$ be given by
(\ref{axial_psi_ab}), (\ref{psi_a}), and (\ref{omega}). If
$|x|\lambda\in E^\infty(B_a(i))$ then $|x|^8\psi_{ab}\psi^{ab}\in
E^\infty(B_a(i))$.
\end{theorem}

The purpose of the condition $|x|^8\psi_{ab}\psi^{ab}\in
E^\infty(B_a(i))$ is to ensure that the solutions to the Hamiltonian
constraints are free of $\ln|x|$ terms. This is the content of the following

\begin{theorem}
\label{existence_theta}
Let $h_{ab}$ be a metric that on $\mathcal{B}_a(i)\in S$
is of the form (\ref{non_smooth}) with $x^ah^{(II)}_{ab}=0$, which
is furthermore smooth on $\widetilde{S}=S\setminus\{i\}$. Let
$\psi_{ab}$ be given by equations (\ref{axial_psi_ab}), (\ref{psi_a}),
and (\ref{omega}). Then there exists a solution $\vartheta$ to
(\ref{Hamiltonian}) which is positive, satisfies the boundary
conditions (\ref{boundary}), and on $B_a(i)$ has the form
\begin{equation}
\vartheta=\frac{\hat{\vartheta}}{|x|}, \quad \hat{\vartheta}\in E^\infty(\mathcal{B}_a(i)), \quad \hat{\vartheta}(i)=1.
\end{equation}
\end{theorem}

\section{The structure equations on $\mathcal{S}$}
\label{section:structure}

Let $\tau^\mu=\sqrt{2}e_0^\mu$, where $e^\mu_0$ is the future $g$-unit
normal of $\mathcal{S}$, and let $SU(\mathcal{S})$ be the bundle of
space-spinors associated with $\tau^{AA'}$, the spinorial counterpart of
$\tau^\mu$. In part I, a certain submanifold $\mathcal{C}_a\subset
SU(\mathcal{S})$ given by
\begin{equation}
\mathcal{C}_a=\bigg\{ \delta(\rho,t)\in SU(\mathcal{S}) \; | \; |\rho|<a, \; t\in SU(2,\Complex) \bigg\},
\end{equation} 
was defined. The manifold $\mathcal{C}_a$ inherits from $SU(\mathcal{S})$ a $\mathfrak{su}(2,\Complex)$-valued \emph{connection form}
$\check{\omega}^A_{\phantom{A}B}$ compatible with the metric
$h_{\alpha\beta}$ and 1-form $\sigma^{AB}$, the \emph{solder form}, such that
\begin{equation}
\label{metric}
h\equiv h_{\alpha\beta} dx^\alpha \otimes dx^\beta = h_{ABCD} \sigma^{AB}\otimes \sigma^{CD},
\end{equation}
where 
\begin{equation}
\label{solder_decomposition}
\sigma^{AB}=\sigma^{AB}_1 d\rho+\sigma^{AB}_+\alpha^+ +\sigma^{AB}_-\alpha^-.
\end{equation}

Let $c_{CD}$ be the vector fields dual to the forms $\sigma^{AB}$, in the sense that
\begin{equation}
\label{c_condition1}
\langle \sigma^{AB}, c_{CD} \rangle=h^{AB}_{\phantom{AB}CD}, \quad c_{CD}=c^1_{CD}\partial_\rho+c^+_{CD}X_++c^-_{CD}X_-,
\end{equation}
where
\begin{equation}
\label{c_condition2}
c^1_{AB}=x_{AB}, \quad c^+_{AB}=\frac{1}{\rho}z_{AB}+\check{c}^+_{AB}, \quad c^-_{AB}=\frac{1}{\rho}y_{AB}+\check{c}_{AB}^-.
\end{equation}
General considerations similar to those in \cite{Fri98a} require that,
\[
\check{c}^\pm_{AB}= Y^\pm y_{AB}+Z^\pm z_{AB},
\]

\bigskip
The relation (\ref{metric}) shows that a knowledge of the conformal
metric $h_{\alpha\beta}$ is equivalent to a knowledge of the solder
form $\sigma^{AB}$, and this, in turn, is equivalent to a knowledge of
the frame vector field $c_{AB}$. Hence, and because of the systematic
use of space spinors in this work, \emph{we shall regard the regular
parts, $\check{c}^\pm_{AB}$, of the frame coefficients, $c^\pm_{AB}$,
as encoding the non-conformal flatness of the conformal metric
$h_{\alpha\beta}$}.

Now, assuming that the conformal 3-metric, $h_{ab}$ is of the form
(\ref{non_smooth}), it follows from the relations (\ref{metric})-(\ref{c_condition2}) and a ``bookkeeping'' argument that
\begin{equation}
\check{c}^\pm_{AB}=(\check{c}^\pm)^{(I)}_{AB}+\rho(\check{c}^\pm)^{(II)}_{AB},
\end{equation}
where $(\check{c}^\pm)^{(I)}_{AB}=\O(\rho^2)$ and
$(\check{c}^\pm)^{(II)}_{AB}=\O(\rho)$ are analytical spinorial fields
on $\mathcal{C}_a$. More precisely, in order to have a class of
initial data sets large enough to contain stationary data we make the
following assumption:

\begin{assumption}
\label{nonsmoothness_frame}
The frame fields $c_{AB}$ on $\mathcal{C}_a$ are
such that
\[
\check{c}^\pm_{AB}=(\check{c}^\pm)^{(I)}_{AB}+\rho(\check{c}^\pm)^{(II)}_{AB}
\]
where $(\check{c}^\pm)^{(I)}_{AB}=\O(\rho^2)$ and $(\check{c}^\pm)^{(II)}_{AB}=\O(\rho)$ are analytical spinorial fields on $\mathcal{C}_a$.
\end{assumption}

\textbf{Remarks.} The spin weights of the diverse components of
$c^\pm_{AB}$ imply under assumption \ref{nonsmoothness_frame} the
following expansion Ans\"atze:

\begin{subequations}
\begin{eqnarray}
&& Y^+= \sum_{p=2}^\infty \sum_{q=2}^p \sum_{m=0}^{2q} Y^+_{p;2q,m} T_{2q\phantom{m}q+2}^{\phantom{2q}m}\rho^p, \quad  Y^-= \sum_{p=2}^\infty \sum_{q=0}^p \sum_{m=0}^{2q} Y^-_{p;2q,m} T_{2q\phantom{m}q}^{\phantom{2q}m}\rho^p, \label{Ansatz_Y}\\
&& Z^+= \sum_{p=2}^\infty \sum_{q=2}^p \sum_{m=0}^{2q} Z^+_{p;2q,m} T_{2q\phantom{m}q-2}^{\phantom{2q}m}\rho^p,  \quad Z^-= \sum_{p=2}^\infty \sum_{q=0}^p \sum_{m=0}^{2q} Z^-_{p;2q,m} T_{2q\phantom{m}q}^{\phantom{2q}m}\rho^p, \label{Ansatz_Z}
\end{eqnarray}
\end{subequations}
where the coefficients $Y^\pm_{p;2q,m}$ and $Z^\pm_{p;2q,m}$ are
complex numbers satisfying the reality conditions
\begin{equation}
\label{Ansatz_reality}
Z^-_{p;2q,m}=(-1)^{m+q}\overline{Y}^+_{p;2q,2q-m}, \quad Y^-_{p;2q,m}=(-1)^{m+q}\overline{Z}^+_{p;2q,2q-m}.
\end{equation}
Note that, for example
\begin{eqnarray*}
&&\hspace{-5mm}Y^-=\left(\frac{1}{2!}\left[Y^-_{2;0,0}\TT{0}{0}{0} +\sum_{m=0}^4Y^-_{2;4,m}\TT{4}{m}{2}\right]\rho^2+\frac{1}{3!}\left[\sum_{m=0}^2 Y^-_{3;2,m}\TT{2}{m}{1} + \sum_{m=0}^6 Y^-_{3;6,m}\TT{6}{m}{3}\right]\rho^3+\cdots\right) \\
&& \phantom{XXXXXX}+\rho \left( \left[ \sum_{m=0}^2Y^-_{2;2,m}\TT{2}{m}{1}\right]\rho +\left[ Y^-_{3;0,0}\TT{0}{0}{0} +\sum_{m=0}^4 Y^-_{3,4,m}\TT{4}{m}{2} \right]\rho^2 +\cdots \right),
\end{eqnarray*}
so that the analytical function contained within the second pair of
round brackets is responsible for the non-analyticity of the frame
vectors ---and of that of the conformal metric. Similar arguments can
be made with the remaining coefficients $Y^+$, $Z^+$ and $Z^-$. More
precisely, one has the following

\begin{lemma}
If the frame fields satisfy assumption \ref{nonsmoothness_frame}, then
in $\mathcal{B}_a(i)$ the conformal metric is of the form
\[
h_{ab}=h^{(I)}_{ab}+|x| h^{(II)}_{ab},
\]
with $h^{(I)}_{ab}$ and $h^{(II)}_{ab}$ analytic.
\end{lemma}
Thus, assumption \ref{nonsmoothness_frame} implies a conformal metric
which is rougher at infinity than that of the Kerr data discussed by
S. Dain in \cite{Dai01c}.

Finally, it is noted that under the assumption of axial symmetry (assumption 0)
the Ans\"{a}tze (\ref{Ansatz_Y})-(\ref{Ansatz_Z}) simplify to
\begin{subequations}
\begin{eqnarray}
&& Y^+= \sum_{p=2}^\infty \sum_{q=2}^p Y^+_{p;2q,q} T_{2q\phantom{q}q+2}^{\phantom{2q}q}\rho^p, \quad  Y^-= \sum_{p=2}^\infty \sum_{q=0}^p Y^-_{p;2q,q} T_{2q\phantom{q}q}^{\phantom{2q}q}\rho^p, \\
&& Z^+= \sum_{p=2}^\infty \sum_{q=2}^p Z^+_{p;2q,q} T_{2q\phantom{q}q-2}^{\phantom{2q}q}\rho^p,  \quad Z^-= \sum_{p=2}^\infty \sum_{q=0}^p  Z^-_{p;2q,q} T_{2q\phantom{q}q}^{\phantom{2q}q}\rho^p,
\end{eqnarray}
\end{subequations}
while the reality condition (\ref{Ansatz_reality}) becomes
\begin{equation}
Z^-_{p;2q,q}=\overline{Y}^+_{p;2q,q}, \quad Y^-_{p;2q,q}=\overline{Z}^+_{p;2q,q}.
\end{equation}

\subsection{Expansions for the connection coefficients}
Now, recall that the contraction of the connection form
$\check{\omega}^A_{\phantom{A}B}$ with the frame $c_{AB}$ defines the
connection coefficients
$\gamma_{CDAB}$. More precisely, we write
\begin{equation}
\gamma_{CDAB}\equiv \langle \check{\omega}^E_{\phantom{A}B},c_{CD}\rangle \epsilon_{EA}= \frac{1}{2\rho}(\epsilon_{AC}x_{BD}+\epsilon_{BD}x_{AC})+\check{\gamma}_{CDAB}=\frac{1}{\rho}\gamma_{CDAB}^{*}+\check{\gamma}_{CDAB},
\end{equation}   
where 
\begin{equation}
\check{\gamma}_{01CD}=0.
\end{equation}
Moreover, it can be checked that
\begin{equation}
\check{\gamma}_{1100}=-\check{\gamma}_{0011}, \quad \check{\gamma}_{0010}=\check{\gamma}_{0001}, \quad \check{\gamma}_{1110}=\check{\gamma}_{1101}.
\end{equation}

As seen in part I, the regular part of the connection coefficients,
$\check{\gamma}_{CD\phantom{A}B}^{\phantom{CD}A}$ vanishes if the
conformal metric is flat. However, if the conformal metric satsifies assumption \ref{nonsmoothness_frame}, then the connection coefficients depend on the frame coefficients $\check{c}^\pm_{AB}$. In order to calculate the connection coefficients we make use of the commutator equations
\begin{equation}
[c_{AB},c_{CD}]=2\bigg( \gamma_{AB\phantom{E}(C}^{\phantom{AB}E}\epsilon_{D)}^{\phantom{D)}F}-\gamma_{CD\phantom{E}(A}^{\phantom{AB}E}\epsilon_{B)}^{\phantom{D)}F}\bigg)c_{EF}+\left(c^-_{AB}c^+_{CD}-c^-_{CD}c^+_{AB}\right)X.
\end{equation}
The latter imply the radial equations,
\begin{subequations}
\begin{eqnarray}
&&\frac{1}{\sqrt{2}}\partial_\rho(\rho\check{c}^+_{AA})=\check{\gamma}_{AA00}(\rho\check{c}^+_{11})-\check{\gamma}_{AA11}(\rho\check{c}^+_{00})-\frac{1}{\sqrt{2}}\check{\gamma}_{AA11},\\
&& \frac{1}{\sqrt{2}}\partial_\rho(\rho\check{c}^-_{AA})=\check{\gamma}_{AA00}(\rho\check{c}^-_{11})-\check{\gamma}_{AA11}(\rho\check{c}^-_{00})-\frac{1}{\sqrt{2}}\check{\gamma}_{AA00}.
\end{eqnarray}
\end{subequations}

The latter equations can be used to obtain expansions for the
connection coefficients $\check{\gamma}_{0000}$,
$\check{\gamma}_{0011}=-\check{\gamma}_{1100}$ and
$\check{\gamma}_{1111}$ in terms of the coefficients in the expansions
of $Y^+$, $Y^-$, and $Z^-$. They also show that  the frame component $Z^+$
is not independent, but can also be expressed in terms of $Y^+$,
$Y^-$, and $Z^-$. Furthermore, one has the following

\begin{lemma}
\label{lemma_connection}
Under assumption \ref{nonsmoothness_frame} it follows that the
connection coefficents $\gamma_{0000}$, $\gamma_{0011}$,
$\gamma_{1100}$ and $\gamma_{1111}$ are such that
\begin{equation}
\check{\gamma}_{ABCD}= (\check{\gamma}_{ABCD})^{(I)}+\rho(\check{\gamma}_{ABCD})^{(II)},
\end{equation}
where $(\check{\gamma}_{ABCD})^{(I)}=\O(\rho^2)$ and
$(\check{\gamma}_{ABCD})^{(II)}=\O(\rho)$ are analytic fields on
$\mathcal{C}_a$. Furthermore, the reality condition
$(\gamma_{ABCD})^+=-\gamma_{ABCD}$ is automatically satisfied if the
conditions \ref{Ansatz_reality} hold.
\end{lemma}

The proof of this lemma follows again from a ``bookkeeping''
argument. Note that the connection coefficients $\gamma_{0001}$ and
$\gamma_{1110}$ are not fixed by the procedure we have discussed
above. We shall come back to them later.

\subsection{The curvature spinors}
In order to determine expansions for diverse components of the
curvature we shall make use of the structure equations. Evaluating the
second structure equation ---equation (36b) in part I --- on
$c_{CD}\wedge c_{EF}$ one obtains the following radial equations
\begin{subequations}
\begin{eqnarray}
&&\frac{1}{\sqrt{2}} \partial_\rho\check{\gamma}_{00AB}+\frac{1}{\rho}\left\{ \check{\gamma}_{0000}z_{AB}-\check{\gamma}_{0011}y_{AB}+\frac{1}{\sqrt{2}}\check{\gamma}_{00AB}\right\} \nonumber \\
&& \phantom{XXXXXXXX}= \check{\gamma}_{0000}\check{\gamma}_{11AB}-\check{\gamma}_{0011}\check{\gamma}_{00AB}-\frac{1}{2}s_{AB00}-\frac{1}{6\sqrt{2}}ry_{AB}, \label{structure_1} \\
&&\frac{1}{\sqrt{2}} \partial_\rho\check{\gamma}_{11AB}+\frac{1}{\rho}\left\{ \check{\gamma}_{1100}z_{AB}-\check{\gamma}_{1111}y_{AB}+\frac{1}{\sqrt{2}}\check{\gamma}_{11AB}\right\}  \nonumber\\
&& \phantom{XXXXXXXX}= \check{\gamma}_{1100}\check{\gamma}_{11AB}-\check{\gamma}_{1111}\check{\gamma}_{00AB}-\frac{1}{2}s_{AB11}-\frac{1}{6\sqrt{2}}rz_{AB} \label{structure_2},
\end{eqnarray}
\end{subequations}
where $s_{ABCD}$ denotes the spinorial representation of the tracefree
part of the Ricci tensor of $h_{\alpha\beta}$, that is,
\begin{equation}
\label{s_definition}
s_{ABCD}=\sigma^i_{AB}\sigma^j_{CD}e^\alpha_i e^\beta_j \bigg( r_{\alpha\beta}-\frac{1}{3}r h_{\alpha\beta} \bigg),
\end{equation}
where $r_{\alpha\beta}$ is the Ricci tensor of $h_{\alpha\beta}$ and
$r=h^{\alpha\beta}r_{\alpha\beta}$. Because of its symmetries, one can write
\begin{equation}
\label{s_components}
s_{ABCD}=s_0\epsilon^0_{ABCD}+s_1\epsilon^1_{ABCD}+s_2\epsilon^2_{ABCD}+s_3\epsilon^3_{ABCD}+s_4\epsilon^4_{ABCD}.
\end{equation}
The equations (\ref{structure_1}) and (\ref{structure_2}) have been
used to obtain expansions for the components of $s_{ABCD}$ in terms of
$Y^+$, $Y^-$, and $Z^-$. More precisely, one has the following

\begin{lemma}
Under assumption \ref{nonsmoothness_frame} one has that the components
of the Ricci spinor are of the form
\[
s_j=s_j^{(I)}+\frac{1}{\rho}s_j^{(II)}
\]
where $s_j^{(I)}=\O(\rho)$ and $s_j^{(II)}=\O(\rho^2)$ are analytic
fields on $\mathcal{C}_a$. The reality condition
$(s_{ABCD})^+=s_{ABCD}$ is satisfied if \ref{Ansatz_reality} holds.
\end{lemma}

\bigskip
 The tracefree part of the Ricci tensor and the Ricci
curvature scalar are not independent from each other, but related via
the Bianchi identity
\[
D^{AB}s_{ABCD}=\frac{1}{6}D_{CD}r.
\]
The latter identity implies three equations which can be used to write
the Ricci scalar $r$ and the connection coefficients $\gamma_{0001}$
and $\gamma_{1110}$ in terms of the frame freely specifiable data
---i.e. $Y^+$, $Y^-$, and $Z^-$. It follows that

\begin{lemma}
Under assumption 2 it follows that the Ricci scalar is of the form
\[
r=r^{(I)}+\frac{1}{\rho}r^{(II)}
\]
where $r^{(I)}=\O(\rho)$ and $r^{(II)}=\O(\rho^2)$ are analytic fields
on $\mathcal{C}_a$. Furthermore, the components $\gamma_{0001}$ and
$\gamma_{1110}$ of the connection are such that
\[
\check{\gamma}_{ABCD}= (\check{\gamma}_{ABCD})^{(I)}+\rho(\check{\gamma}_{ABCD})^{(II)},
\]
where $(\check{\gamma}_{ABCD})^{(I)}=\O(\rho^2)$ and
$(\check{\gamma}_{ABCD})^{(II)}=\O(\rho)$ are analytic fields on
$\mathcal{C}_a$. Furthermore, the reality condition
$(\gamma_{ABCD})^+=-\gamma_{ABCD}$ is automatically satisfied if the
conditions \ref{Ansatz_reality} hold. 
\end{lemma}

\section{Fixing the conformal gauge on $\mathcal{S}$}
\label{section:cngauge}
In our set up, the functions $Y^+$, $Y^-$, and $Z^-$ determining the
frame ---via the relations (\ref{c_condition1}) and
(\ref{c_condition2})--- are used to specify the part of the free data
which is more commonly encoded in the metric. In their expansions,
these functions contain terms which are pure gauge. The reason for
this is the following: given $\vartheta$ and $\psi_{\alpha\beta}$
solutions of the constraint equations (\ref{Hamiltonian}) and
(\ref{momentum}) and a positive function, $\phi$, on $\mathcal{S}$, it
is well known that the transition
\[
h_{\alpha\beta}\mapsto \phi^4h_{\alpha\beta}, \quad \psi_{\alpha\beta}\mapsto \phi^{-2}\psi_{\alpha\beta}, \quad \vartheta \mapsto \phi^{-1}\vartheta, \quad \chi_{\alpha\beta}\mapsto\phi^2\chi_{\alpha\beta}
\]
yields another solution to the Einstein constraint equations with the
 same physical data. In order to fix this conformal freedom we shall
 use a certain gauge based on spacelike conformal geodesics starting
 at $i$. We shall refer to this gauge as the \emph{cn-gauge}.

The \emph{cn-gauge} is defined as follows: consider the 3-dimensional
conformal geodesic equations,
\begin{eqnarray}
&& \dot{x}^\beta\nabla_\beta \dot{x}^\alpha = -2 b_\beta \dot{x}^\beta \dot{x}^\alpha+
\dot{x}_\beta \dot{x}^\beta b^\alpha,  \label{3d_cg_1}\\
&& \dot{x}^\beta\nabla_\beta b_\alpha = b_\beta \dot{x}^\beta b_\alpha
-\frac{1}{2} b_\beta b^\beta \dot{x}_\alpha + \left(s_{\alpha\beta} +
  \frac{1}{12} r h_{\alpha\beta}\right)\dot{x}^\beta, \label{3d_cg_2}
\end{eqnarray}
where $x^\alpha(t)$ is a curve on $\mathcal{S}$, and $b^\alpha$ an
associated 3-dimensional 1-form. We supplement the conformal equations with the initial conditions,
\begin{equation}
\label{3d_cg_initialdata}
x(0)=i, \qquad \dot{x}_{\beta}\dot{x}^\beta=-1, \qquad b(0)=0.
\end{equation}
If $a$ is chosen small enough, there exists a unique solution to these
equations on $\mathcal{B}_a(i)$. Furthermore, there exists in
$\mathcal{B}_a(i)$ an unique conformal rescaling such that,
\begin{equation}
\label{cn_gauge}
b_{\beta}\dot{x}^\beta=0 \qquad \mbox{ on } \mathcal{B}_a(i),
\end{equation}
can always be found. 

\begin{definition}
A metric in the conformal class for which the
condition (\ref{cn_gauge}) is satisfied along the solutions of the
3-dimensional conformal geodesic equations (\ref{3d_cg_1}) and
(\ref{3d_cg_2}) will be said to be in the \emph{cn-gauge}.
\end{definition}

It is noticed that for conformally flat data, if $\dot{x}$ is the
tangent to (standard) geodesics starting at $i$ with
$\dot{x}_{\beta}\dot{x}^\beta=-1$, and if one requires $b\equiv 0$ in
$\mathcal{B}_a(i)$, one is automatically in the cn-gauge.

Let $\dot{X}_{AB}$ and $B_{AB}$ be, respectively, the space spinors
corresponding to the 3-dimensional vector $\dot{x}^\alpha$ and the 1-form
$b_\alpha$. These can be written as,
\begin{equation}
\dot{X}_{AB}=\dot{X}_x x_{AB} + \dot{X}_y y_{AB} + \dot{X}_z z_{AB}, \qquad B_{AB}=B_x x_{AB}+B_y y_{AB} +B_z z_{AB}.
\end{equation} 
Using these two spinors and the spatial Infeld symbols, one can produce
a spinorial version of the conformal geodesic
equations (\ref{3d_cg_1}) and (\ref{3d_cg_2}).

Following our overall strategy, one can attempt tosolve the
3-dimensional conformal geodesic equations together with the cn-gauge
assumption by means of formal expansions. Spin weight considerations
plus the initial data for the conformal geodesics show that
\begin{equation}
 \dot{X}_x=1+\O(\rho), \quad \dot{X}_y=\O(\rho), \quad \dot{X}_z=\O(\rho).
\end{equation}
Similarly, for the components of the spinor $B_{AB}$ one has that
\begin{equation}
B_x=\O(\rho), \quad B_y=\O(\rho), \quad B_z=\O(\rho).
\end{equation}
Once the conformal geodesic equations have been solved, requiring the
cn-gauge condition
$\dot{x}^{\alpha}b_{\alpha}=\epsilon^{AC}\epsilon^{BD}\dot{X}_{CD}B_{AB}=0$
renders further relations between the coefficients in the expansions
of the frame components. Indeed, an explicit calculation with {\tt
Maple V} shows that

\begin{lemma}
Under assumption \ref{nonsmoothness_frame} the cn-gauges imply that
the coefficients $Y^-_{p;2q,m}$ with $2 \leq p \leq 7$, $0 \leq q \leq
p$, $0 \leq m \leq 2q$ are determined by the algebraic combinations of
the coefficients $Y^+_{p;2q,m}$ and $Z^-_{p;2q,m}$ with $2 \leq p \leq
7$, $0 \leq q \leq p$, $0 \leq m \leq 2q$ of $Y^+$ and $Z^-$. The only
exceptions are the coefficients $Y^-_{3;2,m}$, $m=0,1,2$ and
$Y^-_{3;0,0}$ which remain freely specifiable.
\end{lemma}

The full details, being not very illuminating, shall not be given
here. Further consequences of the cn-gauge are that in the
decompositions
\begin{equation}
r=r^{(I)}+\frac{1}{\rho}r^{(II)}, \quad s_j=s^{(I)}_j+\frac{1}{\rho}s^{(II)}_j,
\end{equation}
with $j=0,\ldots,4$ the ``smooth parts'' are of the form
\begin{equation}
r^{(I)}=\O(\rho^2), \quad s^{(I)}_{j}=\O(\rho^2),
\end{equation}
instead of being $\O(\rho)$ ---cfr. with lemma 4.7 in
\cite{Fri98a}. The ``non-smooth parts'' do not change order. It may
also be of interest to note that if the cn-gauge condition holds then
\begin{equation}
\dot{X}_x=\O(\rho^3), \quad \dot{X}_y=\O(\rho^3), \quad \dot{X}_z=\O(\rho^3),
\end{equation}
and
\begin{equation}
B_x=\O(\rho^5), \quad B_y=\O(\rho^2), \quad B_z=\O(\rho^2).
\end{equation}

\section{Expansions of the momentum constraint}

We are now faced with the problem of choosing a class of solutions of the
momentum constraint, equation (\ref{momentum}), which is large enough
to encompass the second fundamental forms of stationary solutions with
non-vanishing ADM angular momentum. In agreement with the existence
results given in section \ref{section:existence}, we shall further
require that the solutions satisfy the regularity condition
$\psi_{\alpha\beta}\psi^{\alpha\beta}\in
E^\infty(\mathcal{B}_a(i))$. The latter requirement is essentially
satisfied if one considers solutions to the momentum constraint
without linear momentum. Our selected class of solutions will be, in
turn, lifted to the manifold $\mathcal{C}_a$, where formal expansions
in terms of powers of $\rho$ will be calculated.

The results by S. Dain discussed in section 2 strongly use the fact
that the initial data possesses axial symmetry to explicitly calculate
---via a ``potential'', $\lambda$--- solutions to the momentum
constraint (\ref{momentum}). However, with future applications in
mind, we are interested in a procedure that can be easily generalised
to a non-axially symmetric setting. 

For the time being let us write
\begin{equation}
\psi_{\alpha\beta}=\psi_{\alpha\beta}^\prime + \widehat{\psi}_{\alpha\beta},
\end{equation}
where both $\psi_{\alpha\beta}^\prime$ and
$\widehat{\psi}_{\alpha\beta}$ are symmetric, tracefree tensors. The
former is a ``seed'' tensor and the latter an unknown to be determined
upon substitution in the momentum constraint. The standard strategy at
this point is to write $\widehat{\psi}_{\alpha\beta}$ in terms of
the Killing operator of a 1-form, $v_\alpha$ ---the York
splitting. For reasons to be explained in section 7, we will not
proceed in this way, but will solve directly for $\psi_{\alpha\beta}$
---see, however, the remarks in subsection \ref{remarks}.

In order to make contact with the conformally flat case studied in
part I, we shall consider as our ``seed'' tracefree tensor
$\psi_{\alpha\beta}^\prime$ the solution to the conformally flat
momentum constraint used in part I ---see assumption 3 in
\cite{Val04e}. As in part I, let
\begin{subequations}
\begin{eqnarray} \label{psi A}
&& \psi_2^A=-\frac{A}{\rho^3}T_{0\phantom{0}0}^{\phantom{0}0}, \\
&& \psi_0^A=\psi_1^A=\psi_3^A=\psi_4^A=0,
\end{eqnarray}
\end{subequations}
\begin{subequations}
\begin{eqnarray}
&&\psi_1^Q=\frac{3}{\rho^2}(Q_2+iQ_1)T_{2\phantom{0}0}^{\phantom{2}0}-\frac{3\sqrt{2}}{\rho^2}Q_3T_{2\phantom{1}0}^{\phantom{2}1}-\frac{3}{\rho^2}(Q_2-iQ_1)T_{2\phantom{2}0}^{\phantom{2}2}, \label{psi Q1}\\
&&\psi_2^Q=\frac{9\sqrt{2}}{2\rho^2}(Q_2+iQ_1)T_{2\phantom{0}1}^{\phantom{2}0}-\frac{9}{\rho^2}Q_3T_{2\phantom{1}1}^{\phantom{2}1}-\frac{9\sqrt{2}}{2\rho^2}(Q_2-iQ_1)T_{2\phantom{2}1}^{\phantom{2}2} \label{psi Q2}\\
&&\psi_3^Q=-\frac{3}{\rho^2}(Q_2-iQ_1)T_{2\phantom{2}2}^{\phantom{2}2}-\frac{3\sqrt{2}}{\rho^2}Q_3T_{2\phantom{1}2}^{\phantom{2}1}+\frac{3}{\rho^2}(Q_2+iQ_1)T_{2\phantom{2}2}^{\phantom{2}0}, \label{psi Q3} \\
&& \psi^Q_0=\psi^Q_4=0, \label{psi Q4}
\end{eqnarray}
\end{subequations}
\begin{subequations}
\begin{eqnarray}
&&\psi_1^J=\frac{6}{\rho^3}(-J_1+iJ_2)\TT{2}{0}{0}+\frac{6\sqrt{2}}{\rho^3}iJ_3\TT{2}{1}{0}-\frac{6}{\rho^3}(J_1+iJ_2)\TT{2}{2}{0}, \label{psi J1}\\
&&\psi_2^J=0 \label{psi J2}\\
&&\psi_3^J=\frac{6}{\rho^3}(J_1+iJ_2)\TT{2}{2}{2}-\frac{6\sqrt{2}}{\rho^3}iJ_3\TT{2}{1}{2}-\frac{6}{\rho^3}(-J_1+iJ_2)\TT{2}{0}{2}, \label{psi J3} \\
&& \psi^J_0=\psi^J_4=0, \label{psi J4}
\end{eqnarray}
\end{subequations}
were $A$, $J_1$, $J_2$, $J_3$, $Q_1$, $Q_2$,
$Q_3\in \Real$. Recall that assumption 0 ---axial symmetry--- requires
$Q_1=Q_2=J_1=J_2=0$. Furthermore, we write
\begin{equation} \label{psi_lambda01234}
\psi_j^\lambda =\frac{1}{\rho^4} \sum_{n=2}^\infty \sum_{q=2}^n  L_{j,n-4;2q,q} \TT{2q}{q}{q-2+j} \rho^n, \quad j=0,\ldots,4
\end{equation}
where
\begin{subequations}
\begin{eqnarray}
&&\hspace{-2cm} L_{1,n-4;2q,q}=\frac{\left( (L_{4,n-4;2q,q}-L_{0,n-4;2q,q})(q+1)q+4L_{0,n-4;2q,q}(n+3)^2 \right)}{(n+3)(2(n+4)(n+2)-(q+2)(q-1))}\sqrt{(q+2)(q-1)}, \label{K1}\\
&&\hspace{-2cm} L_{2,n-4;2q,q}= \frac{3(L_{0,n-4;2q,q}+L_{4,n-4;2q,q})}{2(n+3)^2-q(q+1)}\sqrt{(q+2)(q+1)q(q-1)},\label{K2} \\
&&\hspace{-2cm} L_{3,n-4;2q,q}= \frac{\left( (L_{0,n-4;2q,q}-L_{4,n-4;2q,q})(q+1)q+4L_{4,n-4;2q,q}(n+3)^2 \right)}{(n+3)(2(n+4)(n+2)-(q+2)(q-1))}\sqrt{(q+2)(q-1)}, \label{K3}
\end{eqnarray}
\end{subequations}
and $n=2,3,4,\ldots$, axial symmetry (assumption 0) already being
assumed here. The coefficents $L_{0;2q,q}$ and $L_{4;2q,q}$ are
arbitrary complex numbers satisfying the reality condition
\begin{equation}
L_{0;2q,q}=\overline{L}_{4;2q,q}.
\end{equation}
We shall make the following
\begin{assumption}
\label{2nd_seed_assumption}
The totally symmetric spinor $\psi_{ABCD}^\prime$ associated with the
``seed'' tracefree tensor $\psi_{\alpha\beta}^\prime$ in the York
splitting (\ref{york_splitting}) is assumed to be of the form
\begin{equation}
\psi_{ABCD}^\prime=\psi^A_{ABCD}+\psi^Q_{ABCD}+\psi^J_{ABCD}+\psi^\lambda_{ABCD},
\end{equation}
where
\begin{equation}
\psi^i_{ABCD}=\psi^i_0\epsilon^0_{ABCD}+\psi^i_1\epsilon^1_{ABCD}+\psi^i_2\epsilon^2_{ABCD}+\psi^i_3\epsilon^3_{ABCD}+\psi^i_4\epsilon^4_{ABCD},
\end{equation}
and $i$ denotes any of the labels $A$, $Q$, $J$, $\lambda$.
\end{assumption}

\bigskip
The symmetric, tracefree tensor $\widehat{\psi}_{\alpha\beta}$ has a spinorial counterpart $\widehat{\psi}_{ABCD}$ of the form
\begin{equation}
\widehat{\psi}_{ABCD}=\widehat{\psi}^i_0\epsilon^0_{ABCD}+\widehat{\psi}^i_1\epsilon^1_{ABCD}+\widehat{\psi}^i_2\epsilon^2_{ABCD}+\widehat{\psi}^i_3\epsilon^3_{ABCD}+\widehat{\psi}^i_4\epsilon^4_{ABCD}.
\end{equation}
Under assumption \ref{2nd_seed_assumption}, one has that
\begin{equation}
D^{AB}\psi^\prime_{ABCD}=\O(\rho).
\end{equation}
Consistent with our general strategy of finding solutions to the
constraint equations which are expandable in powers of $\rho$ near
infinity, we put forward the Ansatz
\begin{equation}
\widehat{\psi}_{j}=\sum^\infty_{p=2}\sum^p_{q=|2-l|}\sum^{2q}_{m=0}\widehat{\psi}_{j,p;2q,m}\TT{2q}{m}{q-2+j}\rho^p,
\end{equation}
where the coefficients $\widehat{\psi}_{j,p;2q,m}$ satisfy the
required reality conditions ---see part I. Under assumption 0 ---axial symmetry--- our Ansatz simplifies to
\begin{equation}
\label{psihat}
\widehat{\psi}_{j}=\sum^\infty_{p=2}\sum^p_{q=|2-l|}\widehat{\psi}_{j,p;2q,q}\TT{2q}{q}{q-2+j}\rho^p.
\end{equation}
In this case, substitution into the momentum constraint yields
algebraic equations for the coefficients
$\widehat{\psi}_{j,p;2q,q}$. These allow us to write the coefficients
$\widehat{\psi}_{1,p;2q,q}$, $\widehat{\psi}_{2,p;2q,q}$ and
$\widehat{\psi}_{3,p;2q,q}$ in terms of the freely specifiable data
and in terms of the coefficients $\psi_{0,p;2q,q}$ and
$\psi_{4,p;2q,q}$. The latter are related to each other by the reality
condition
\begin{equation}
\widehat{\psi}_{4,p;2q,q}=\overline{\widehat{\psi}}_{0,p;2q,q}.
\end{equation}
The relations discussed in the above paragraph have been explicitly
calculated in the CA system {\tt Maple V} up to order $p=4$ in the
Ansatz (\ref{psihat}).

\subsection{A remark concerning the York splitting} \label{remarks}
For the sake of completeness, it is of interest to analyse what
happens if instead of using the procedure described in the previous
subsection, one fully uses the York splitting. Accordingly, let us
write
\begin{equation}
\label{york_splitting}
\psi_{\alpha\beta}=\psi_{\alpha\beta}^\prime + (\mathscr{L}v)_{\alpha\beta},
\end{equation}
where 
\begin{equation}
(\mathscr{L}v)_{\alpha\beta}=D_\alpha v_\beta + D_\beta v_\alpha -\frac{2}{3}h_{\alpha\beta}D_\gamma v^\gamma
\end{equation}
is the \emph{conformal Killing operator}. Let $v_{AB}$ be the
spinorial counterpart of the 1-form $v_{\alpha}$ on which the
conformal Killing operator in the York splitting ---equation
(\ref{york_splitting}) acts, and write
\begin{equation}
v_{AB}=v_x x_{AB}+v_y y_{AB}+v_z z_{AB},
\end{equation}
where the coefficients $v_x$, $v_y$ and $v_z$ satisfy the reality conditions
\begin{equation}
v_x=\overline{v}_x, \quad v_y=\overline{v}_z.
\end{equation}
Under assumption \ref{2nd_seed_assumption} one has that
\begin{equation}
D^{AB}\psi^\prime_{ABCD}=\O(\rho).
\end{equation}
The latter suggests the following Ansatz for the components of $v_{AB}$:
\begin{subequations}
\begin{eqnarray}
&& v_x=\sum_{p=1}^\infty \sum_{q=0}^p \sum_{m=0}^{2q} v_{x,p;2q,m}\TT{2q}{m}{q}\rho^p, \label{ansatz_vx}\\
&& v_y=\sum_{p=1}^\infty \sum_{q=1}^p \sum_{m=0}^{2q} v_{y,p;2q,m}\TT{2q}{m}{q+1}\rho^p, \label{ansatz_vy} \\
&& v_z=\sum_{p=1}^\infty \sum_{q=1}^p \sum_{m=0}^{2q} v_{z,p;2q,m}\TT{2q}{m}{q-1}\rho^p, \label{ansatz_vz}
\end{eqnarray}
\end{subequations}
where because of the reality conditions one has
\begin{equation}
\label{v_reality}
v_{y,p;2q,m}=(-1)^{q+m+1}\overline{v}_{z,p;2q,m}.
\end{equation}

Under assumption 0 ---axial symmetry--- the above expressions reduce to
\begin{subequations}
\begin{eqnarray}
&& v_x=\sum_{p=1}^\infty \sum_{q=0}^p v_{x,p;2q,q}\TT{2q}{q}{q}\rho^p, \\
&& v_y=\sum_{p=1}^\infty \sum_{q=1}^p v_{y,p;2q,q}\TT{2q}{q}{q+1}\rho^p, \\
&& v_z=\sum_{p=1}^\infty \sum_{q=1}^p v_{z,p;2q,q}\TT{2q}{m}{q-1}\rho^p,
\end{eqnarray}
\end{subequations}
with 
\begin{equation}
v_{y,p;2q,m}=-\overline{v}_{z,p;2q,m}.
\end{equation}

The substitution of this Ansatz in the spinorial counterpart of the
conformal Killing operator $(\mathscr{L}v)_{ABCD}$, and then in turn
in
\[
\label{momentum_spinorial}
D^{AB}\bigg(\psi^\prime_{ABCD}+(\mathscr{L}v)_{ABCD}\bigg)=0,
\]
renders a series of algebraic equations for the coefficients
$v_{x,p;2q,q}$, $v_{y,p;2q,q}$ and $v_{z,p;2q,q}$. Calculations using
{\tt Maple V} reveal that at least for $p=1,\ldots,4$, the
coefficients $v_{x,p;2p,p}$, $v_{y,p;2p,p}$ and $v_{z,p;2p,p}$
---i.e. for the sector $q=p$--- can be written in terms of the freely
specifiable data ---i.e. the coefficients $Y^+_{p;2q,q}$,
$Z^-_{p;2q,q}$, $L_{0,p;2q,q}$, $L_{4,p;2q,q}$, and the constants $A$,
$J$, $Q_3$. Now, for $q<p$ the situation is different. It is found
that for arbitrary specifiable data, there is in general, no solution
to the algebraic equations linking the coefficients $v_{x,p;2q,q}$,
$v_{y,p;2q,q}$ and $v_{z,p;2q,q}$. Nevertheless, conditions for the
existence of solutions can be found. These restrictions on the
specifiable data can be understood by recalling that generic solutions
of the momentum constraint, equation (\ref{momentum}), will have a much
more complicated asymptotic behaviour than the one we have imposed
through the Ans\"atze (\ref{ansatz_vx})-(\ref{ansatz_vz}). One could
have solutions containing logarithmic terms, for
example. Consequently, conditions on the free data are needed in order
to have consistency with the asymptotic behaviour which is being
pursued. The conditions are given in the following

\begin{theorem}
\label{theorem:momentum}
Under assumptions 1, 2 and 3, necessary conditions for the existence
of solutions to the momentum constraint, equation (\ref{momentum}),
which on $\mathcal{C}_a$ are expandable in powers of $\rho$ up to
order $\O(\rho^4)$ are
\[
Z^-_{p;2q,q}=(-1)^{p+q+1}Y^+_{p;2q,q}, \quad p=2,\ldots,5, \;\; q=2,\ldots,p, \;\; m=0,\ldots,2q,
\] 
\end{theorem}

A further peculiarity of solutions to equation
(\ref{momentum_spinorial}) under the Ans\"atze
(\ref{ansatz_vx})-(\ref{ansatz_vz}) is that even if the conditions
given in theorem \ref{theorem:momentum} are satisfied, the
coefficients $v_{x,p;2p,q}$, $v_{y,p;2p,q}$ and $v_{z,p;2p,q}$ are not
fully determined by the algebraic conditions implied by the
substitution of (\ref{ansatz_vx})-(\ref{ansatz_vz}) in equation
(\ref{momentum_spinorial}). More precisely, if the conditions given in
theorem \ref{theorem:momentum} hold then it follows that
\begin{equation}
v_{y,p;2q,q}+(-1)^{p+q+1}v_{z,p;2q,q}=C_{p;2q,q}
\end{equation}
for $p=1,\ldots,4$, $q=1,\ldots,p$ and where $C_{p;2q,q}$ denotes some
expression depending on the freely specifiable data. Recall that
besides, the reality condition (\ref{v_reality}) must also be
satisfied. The reason for this behaviour can be understood as
follows. It is well known that the idea behind the York splitting,
equation (\ref{york_splitting}), is to obtain an elliptic equation for
the 1-form $v_\alpha$ out of the underdetermined system
(\ref{momentum}). Now, solutions to a system of elliptic equations
contain global information about the underlying initial manifold
$\mathcal{S}$. This information is not captured if one attempts to
calculate formal solutions in a neighbourhood $\mathcal{B}_a(i)$ of
infinity. This same feature pervades the approach for solving the
momentum constraint followed in the previous section. Moreover, a
similar phenomenon can be observed when ``solving'' the Hamiltonian
constraint by the same methods ---see in particular the discussion in
part I.

\section{Expansions of the Hamiltonian constraint}
As discussed in part I, the Licnerowicz equation, being a scalar
equation is easily translated into space spinor language. Indeed, one
has
\begin{equation}
\label{licnerowicz_spinorial}
\bigg(D^{AB}D_{AB}-\frac{1}{8}r\bigg)\vartheta=\frac{1}{8}\psi_{ABCD}\psi^{ABCD}\vartheta^{-7}.
\end{equation}
In order to construct formal expansions of the solutions of this
equations on $\mathcal{B}_a(i)$ ---respectively $\mathcal{C}_a$--- we
adopt the local parametrisation
\begin{equation}
\label{theta_parametrisation} 
\vartheta=\frac{U}{|x|}+W.
\end{equation}
The term $U/|x|$ (the Green function) contains information regarding
the local geometry near infinity. In the conformally flat case on has
$U=1$. On the other hand, the term $W$ contains information of a
global nature. For example, $W(i)=m/2$, where $m$ is the ADM mass of
the initial data set 
$(\widetilde{\mathcal{S}},\widetilde{h}_{\alpha\beta},\widetilde{\chi}_{\alpha\beta})$
---the global quantity \emph{par excellence}.

\subsection{The construction of the Green function}
In order to calculate the function $U$, we shall make use of
\emph{Hadamard's parametrix construction} ---see
e.g. \cite{Fri98a,Fri04}. The term $U/|x|$, being the Green's function
of the \emph{Yamabe operator} ---see below---, corresponds to a
solution of the equation
\begin{equation}
\label{Green_equation}
\left(D^\alpha D_\alpha -\frac{1}{8}r\right)\left( \frac{U}{|x|}\right)=4\pi \delta_i,
\end{equation}
where $\delta_i$ is the Dirac distribution centred in $i$. Now,
consider a function $U$ of the form
\begin{equation}
U=\sum^\infty_{p=0}U_p|x|^{2p}, \label{series_U}
\end{equation}
where $U_p=U_p(|x|)$, that is, the coefficients in the Ansatz are still
allowed to have $\rho$-dependence. Under this Ansatz, equation
(\ref{Green_equation}) implies the following hierarchy of ordinary
differential equations:
\begin{eqnarray*}
&& D^\alpha \rho^2 D_\alpha U_0= -\frac{1}{2}(D^\alpha D_\alpha \rho^2 + 6)U_0, \quad U(i)=1, \\
&& D^\alpha \rho^2 D_\alpha U_p =-\frac{1}{2}(D^\alpha D_\alpha \rho^2 +6 -4p)U_p -\frac{1}{2p-1}L_h[U_{p-1}], \quad p=1,2,\dots
\end{eqnarray*}
where 
\[
L_h[f]=(D^\alpha D_\alpha -1/8 r)f,
\]
denotes the Yamabe operator of the metric $h_{\alpha\beta}$ applied to
a smooth function $f$. The solutions to the above equations can be
given recursively by
\begin{eqnarray*}
&& U_0=\mbox{exp}\left( \frac{1}{4} \int_0^\rho (D^\alpha D_\alpha s^2 +6)\frac{\d s}{s} \right), \\
&& U_{p+1}=-\frac{U_0}{(4p-2)\rho^{p+1}} \int_0^\rho \frac{L_h[U_p] s^p}{U_0} \d s, \quad p=0,1,\dots.
\end{eqnarray*}
Now, given a conformal 3-metric $h_{\alpha\beta}$, the above procedure
does not necessarily render a convergent $U$ as given by 
the series (\ref{series_U}). One needs extra assumptions on $h$. For
example, if $h_{\alpha\beta}$ is analytic in a neighbourhood of $i$
then $U$ given by (\ref{series_U}) is also analytic.

Under the more general requirements of assumption
\ref{nonsmoothness_frame} the best one can hope for is a function $U$
of a similar form ---i.e. $U=U^{(1)}+\rho U^{(2)}$. The latter will be
the case only if further conditions on the conformal metric hold, for
some experimentation reveals that free data of the form given by
assumption \ref{nonsmoothness_frame} render coefficients $U_p$ with
logarithmic terms in $|x|$. Direct calculations with {\tt Maple V}
yield the following

\begin{theorem}
\label{regularity_of_U}
Under assumption \ref{nonsmoothness_frame}, the coefficients
$U_0,\ldots,U_5$ have a logarithm-free dependence on $|x|$ if and only
if
\begin{subequations}
\begin{eqnarray*}
&& Z^-_{2;4,2}=-Y^+_{2;4,2}, \\
&& Y^-_{3;0,0}=0, \quad Y^-_{3;2,1}=0, \\
&& Z^-_{3;6,3}=-Y^+_{3;6,3}, \\
&& Z^-_{4;4,2}=-Y^+_{4;4,2}, \quad Z^-_{4;8,4}=-Y^+_{4;8,4}, \\
&& Z^-_{5;6,3}=-Y^+_{5;6,3}, \quad Z^-_{5;10,5}=-Y^+_{5;10,5}.
\end{eqnarray*}
\end{subequations}
If the above conditions hold, then 
\begin{subequations}
\begin{eqnarray*}
&& U_0=1+\O(|x|^4),\\
&& U_1=\O(|x|^2), \\
&& U_2=\O(|x|^2), \\
&& U_3=\O(|x|^0), \\
&& U_4=\O(|x|^{-1}), \\
&& U_5=\O(|x|^{-3}).
\end{eqnarray*}
\end{subequations}
\end{theorem}

\textbf{Remark.} The conditions given in
theorem \ref{regularity_of_U} allow to calculate an expansion for $U$ up
to order $\O(\rho^7)$.

\subsection{Construction of expansions of $W$}
The substitution of the parametrisation (\ref{theta_parametrisation})
in the Hamiltonian constraint, equation (\ref{Hamiltonian}), renders
the following equation for the function $W$ in $\mathcal{C}_a$:
\begin{equation}
\label{eqn_for_W}
\bigg( D^{AB} D_{AB} -\frac{1}{8}r \bigg)W =\frac{1}{8}\psi_{ABCD}\psi^{ABCD}(U/\rho+W)^{-7}.
\end{equation}
Under the assumption of axial symmetry (assumption 0), we shall be
looking for solutions on $\mathcal{C}_a$ of the form
\begin{equation}
\label{Ansatz_W}
W=m/2+\sum^\infty_{p=1} \sum_{q=0}^p \frac{1}{p!}w_{p;2q,q} \TT{2q}{q}{q}\rho^p,\end{equation}
with $w_{p;2q,m}\in \Real$ so that $W$ is a real function on $\mathcal{C}_a$.

As in the conformally flat case discussed in part I, the substitution
of the Ansatz (\ref{Ansatz_W}) into equations (\ref{eqn_for_W}) allows
to determine the coefficients $w_{p;2q,q}$, with $q<p$. The
coefficients $w_{p;2p,p}$ ---like the ADM mass itself--- are, on the
other hand, not fixed by this procedure, and shall remain free in our
construction. Note, however, that in an actual solution of the
constraints, these coefficients are functions of the free data. This
dependence can be analysed, for example, by writing an integral
representation of the solution ---see e.g. the discussion in part I or
in \cite{Dai02a}.

In part I, it was shown that the dipolar terms in the expansions of
$W$ can be, without loss of generality, chosen to be zero. This could
be done because our setup still allowed for a translational
freedom. In our case, due to the use of cn-gauge to fix the conformal
gauge this is in general not possible. Nevertheless, in order to ease the
complexity of our computations we make the following

\begin{assumption}
\label{centre_of_mass}
We shall restrict our assumption to initial data sets such that
the function $W$ in the parametrisation $\vartheta=U/\rho+W$ of the
conformal factor is of the form
\[
W=\frac{m}{2}+\O(|x|^2).
\]
\end{assumption}

Assumption \ref{centre_of_mass} implies that in the expansion
(\ref{Ansatz_W}) one has $w_{1;2,m}=0$ for $m=0,1,2$. For the purposes
of this article the expansions of $W$ have been explicitly calculated
up to order $\O(\rho^7)$.

\section{Regularity conditions on the initial data}

As in part I, let us expand the components of the Weyl spinor $\phi_{ABCD}$ as 
\[
\phi_j=\sum_{p=|2-j|}^\infty\frac{1}{p!}\sum_{q=|2-j|}^p\sum_{m=0}^{2q} a_{j,p;2q,m}\TT{2q}{m}{q-2+j}\rho^p,
\]
where the (complex) coefficients $a_{j,p;2q,m}$ have $\tau$ dependence
---that is, $a_{j,p;2q,m}=a_{j,p;2q,m}(\tau)$. Friedrich ---see
\cite{Fri98a}--- has shown that the solutions to the transport
equations implied by the conformal Einstein equations at spatial
infinity develop a certain type of logarithmic singularity at the sets
where null infinity ``touches'' spatial infinity unless the initial
data is such that
\begin{equation}
\label{regcond_1}
a_{0,p;2p,k}(0)=a_{4,p;2p,k}(0)
\end{equation}
with $p=2,3,\ldots$, and $k=0,\ldots,2q$ holds.

In part I, the regularity condition (\ref{regcond_1}) was rephrased in
terms of the freely specifiable terms available in the case of
conformally flat initial data sets. It essentially implied that
the coefficients $L_{0,p-4;2p,k}$ and $L_{4,p-4;2p,k}$ encoding part
of the higher multipole structure of the conformally flat initial data
sets have to be alternatively real or pure imaginary numbers depending
on the parity of $k$ ---see theorem 3 in \cite{Val04e}.

In the case of initial data of the form being considered here, we have the following

\begin{theorem}
\label{theorem:regularity}
For an initial data set satisfying the assumptions 1-4, the following conditions are equivalent:
\begin{itemize}
\item[(i)] $\phi_{0,p;2p,p}(0)=\phi_{4,p;2p,p}(0)$, $p=2,3,4,5,6$;

\item[(ii)]
\begin{subequations} 
\begin{eqnarray}
&& Y^+_{2;4,2}=\frac{2}{3}\left(L_{4,-2;4,2}+L_{0,-2;4,2}\right), \\
&& Y^+_{3;6,3}=\frac{9}{4}\left(L_{4,-1;6,3}+L_{0,-1;6,3}\right), \\
&& Y^+_{4;8,4}=\frac{96}{5}\left(L_{0,0;8,4}+L_{4,0;8,4}\right), \\
&& Y^+_{5;10,5}=-100\left(L_{0,1;10,5}+L_{4,1;10,5}\right), \\
&& Y^+_{6;12,6}=-\frac{1080}{7}\left(L_{0,2;12,6}+L_{4,2;12,6}\right).
\end{eqnarray}
\end{subequations}
\end{itemize}

\end{theorem}

This result naturally leads to our final assumption.

\begin{assumption}
\label{assumption:regularity}
The second fundamental form satisfies the condition (ii) in theorem
\ref{theorem:regularity}.
\end{assumption}

\section{Solutions to the transport equations and obstructions to the smoothness of null infinity}

We shall now proceed, in a parallel way as it was done in part I, to
discuss the properties of the solutions of the transport equations
implied by the conformal Einstein equations ---see equations (131),
(132) and (133)--- in the case of initial data sets satisfying
assumptions 0-5. For details on the structure of the transport
equations we remit the reader to part I. We just recall that the
transport equations implied by the conformal Einstein equations upon
evaluation on the cylinder at spatial infinity naturally group in two
subsystems. One, called the \emph{v subsystem}, contains the evolution
equations for the frame, connection, and the components of the Ricci
tensor of the Weyl connection. The other one, the so-called
\emph{Bianchi subsystem}, provides equations for the components of the
Weyl tensor. As in references \cite{Val04a,Val04d,Val04e}, we shall be
restricting our attention to the logarithmic terms arising in the components
of the Weyl spinor $\phi_{ABCD}$ ---the reasons for this way of
proceeding have been discussed in the aforementioned references. To
this end we recall that we are decomposing the components of the Weyl
tensor as
\begin{equation}
\phi_j=\sum_{p=|2-j|}^\infty \frac{1}{p!} \phi_j^{(p)}\rho^p,
\end{equation}
where
\begin{equation}
 \phi_j^{(p)}=\sum_{q=|2-j|}^p\sum_{k=0}^{2q} a_{j,p;2q,k}(\tau)\TT{2q}{k}{q-2+j}.
\end{equation}
Retaking a notation already used in part I, we use $\mathcal{Q}(\tau)$
to denote a generic polynomial in $\tau$, while $\mathcal{P}_k(\tau)$
will denote a generic polynomial of degree $k$ in $\tau$ such that
$\mathcal{P}_k(\pm 1)=0$. It will be understood that the
$\mathcal{Q}(\tau)$ and $\mathcal{P}_k(\tau)$ appearing in different
equations and/or components  are also different from each other.

\subsection{Lower order solutions}

Our first result is the following

\begin{theorem}
Under assumptions 0-5, the solutions of both the v and Bianchi
transport equations for $p=0,1,2$ are polynomial in $\tau$. In
addition, the solutions of the v transport equations are polynomial for
$p=3$.
\end{theorem}

A first difference with respect to the cases analysed in paper I and
also in \cite{Val04a,Val04d,FriKan00}, is that extra conditions are
needed to calculate the solutions of the $p=3$ Bianchi transport
subsystem. Namely, one has that

\begin{theorem}
Under assumptions 1-5, the Bianchi transport constraint equations are
satisfied if and only if
\begin{equation} \label{violation_b342}
L_{4,-2;4,2}=-L_{0,-2;4,2}.
\end{equation}
Furthermore, if the latter condition holds then the solutions of the
Bianchi transport equations for $p=3$ are polynomial in $\tau$.
\end{theorem}

\textbf{Remark.} It is noted that the condition (\ref{violation_b342})
arose in paper I as a regularity condition which precluded the
appearance of logarithms associated to the ``highest possible''
harmonics. In the present case (\ref{violation_b342}) implies, in
turn, that
\[
Y^+_{2;4,2}=Y^-_{2;4,2}=Z^+_{2;4,2}=Z^-_{2;4,2}=0,
\]
from where it follows that $\check{c}^\pm_{AB}=\O(\rho^3)$.

\subsection{The solutions at order $p=4$}

Now, let the assumptions 1-5 together with the condition
(\ref{violation_b342}) hold. The CA computations show that the
solutions to the $p=4$ v transport equations are, as it is to be
expected, polynomial in $\tau$. In complete analogy to the $p=3$ case
one has that some extra conditions need to be satisfied in order to be
able to solve the $p=4$ Bianchi transport equations. More precisely,
one has the following

\begin{theorem}
Let assumptions 0-5 together with condition
(\ref{violation_b342}). Then the $p=4$ Bianchi constraint equations
are satisfied if and only if
\begin{equation} \label{violation_b463}
L_{4,-1;6,3}=-L_{0,-1;6,3}.
 \end{equation}
Furthermore the functions $a_{j,4;2q,q}(\tau)$, $q=j,\ldots, 4$ $q\neq 2$ are polynomial in $\tau$, while
\begin{equation}
a_{j,4;4,2}(\tau)=\Upsilon_{4;4,2}\bigg( (1-\tau)^{6-j}\ln(1-\tau)\mathcal{P}_j(\tau) +(1+\tau)^{2+j}\ln(1+\tau)\mathcal{P}_{4-j}(\tau) \bigg) +\mathcal{Q}(\tau),
\end{equation}
where
\begin{equation} \label{obstruction_b442}
\Upsilon_{4;4,2}=Z^-_{3;4,2}-Y^+_{3;4,2}.
\end{equation}

\end{theorem}

\medskip
\textbf{Remark 1.} Again, because of the regularity conditions given
in theorem \ref{theorem:regularity}, if the condition
(\ref{violation_b463}) holds then one has that
\begin{equation}
Y^+_{3;6,3}=Y^-_{3;6,3}=Z^+_{3;6,3}=Z^-_{3;6,3}=0.
\end{equation}

\medskip
\textbf{Remark 2.} The obstruction (\ref{obstruction_b442}) is a
result of the non-conformal flatness of our setting. It did not arise
in the analysis carried out in \cite{Val04d} as the free data was chosen
in such a way that it was automatically satisfied. It is noted that
had we solved the momentum constraint in the way indicated in section
\ref{remarks} then the obstruction (\ref{obstruction_b442}) would not
have arosen ---see theorem \ref{theorem:momentum}. As it will be seen
in the sequel, analogous higher order obstruction would have,
similarly, gone unnoticed.

\subsubsection{Solutions at order $p=5$}

For the solutions of the $p=5$ transport equations one has the following 

\begin{theorem}  
Consider an initial data set satisfying the conditions 0-5. Assume,
additionally, that the conditions (\ref{violation_b342}),
(\ref{violation_b463}) also hold and the obstruction
(\ref{obstruction_b442}) vanishes. Then the solutions to the $p=5$
v transport equations are polynomial in $\tau$ and the transport
equations implied by the Bianchi constraint equations at order $p=5$ 
are satisfied if and only if
\begin{equation} \label{violation_b584}
L_{4,0;8,4}=-L_{0,0;8,4}.
\end{equation}
If the latter holds one has that $a_{j,5;2q,q}(\tau)$, $j=0,\ldots,4$, $q=|2-j|,\ldots,5$ with $q\neq 2,3$ are polynomial in $\tau$ and 
\begin{equation}
a_{j,5;4,2}(\tau)=\Upsilon_{5;4,2}\big((1-\tau)^{7-j}\mathcal{P}_j(\tau)\ln(1-\tau)+(1+\tau)^{3+j}\mathcal{P}_{4-j}(\tau)\ln(1+\tau)\big)+\mathcal{Q}(\tau),
\end{equation}
where
\begin{eqnarray}
&& \Upsilon_{5;4,2}=18w_{2;4,2}m^2+m R_{2;4,2}+\frac{37602}{199}mJ^2-\frac{3099}{199}\sqrt{6}m^2L_{0,-2;4,2} \nonumber \\
&& \phantom{XXXXX}+\frac{1023}{199}\sqrt{6}m\Big(L_{0,-1;4,2}-L_{4,-1;4,2}\Big)+\frac{224}{199}\sqrt{6}\Big(L_{0,0;4,2}-L_{4,0;4,2}\Big). \label{obstruction_b542}
\end{eqnarray}
Furthermore,
\begin{equation}
a_{j,5;6,3}(\tau)=\Upsilon_{5;6,3}\bigg((1-\tau)^{7-j}\ln(1-\tau)\mathcal{P}_{j+1}+ (1+\tau)^{3+j}\ln(1+\tau)\mathcal{P}_{5-j}(\tau)\bigg)+\mathcal{Q}(\tau),
\end{equation}
where
\begin{equation} \label{obstruction_b563}
\Upsilon_{5;6,3}=Z^-_{4;6,3}-Y^+2_{4;6,3}.
\end{equation}
\end{theorem}

\textbf{Remark 1.} The expression (\ref{obstruction_b542}) is the
natural generalisation of the quadrupolar obstructions obtained in
paper I and in \cite{Val04a,Val04d}. In particular, in order to make
the comparison with the results of \cite{Val04d} more transparent we
note that the calculations leading to lemma 4 in section 3.2 show that
\begin{equation}
r_{2;4,2}=16\sqrt{6}Y^+_{3;4,2},
\end{equation}
where
\[
r=r_{2;4,2}\TT{4}{2}{2}\rho^2+\O(\rho^3),
\]
and $r$ is the Ricci scalar of the conformal metric as discussed in
section 3. Thus, in the time symmetric case ---i.e. if
$J=L_{0,-2;4,2}=L_{0,-1;4,2}=L_{4,-1;4,2}=L_{0,0;4,2}=L_{4,0;4,2}=0$---
the quadrupolar obstruction (\ref{obstruction_b542}) reduces to
\begin{equation}
\Upsilon_{5;4,2}=18w_{2;4,2}m^2+m R_{2;4,2}.
\end{equation}

\textbf{Remark 2.} Note that the first three terms in
(\ref{obstruction_b542}) are manifestly real, while the remaining
three are ---due to the reality conditions and the condition
(\ref{violation_b342}) in theorem 7--- pure imaginary. Thus, the
vanishing of the quantity $\Upsilon_{5;4,2}$ implies, on the one hand,
that
\begin{equation}
mw_{2;4,2}=-\frac{8\sqrt{6}}{9}Y^+_{3;4,2}-\frac{2089}{199}J^2, 
\end{equation}
and on the other hand
\begin{equation}
1023m(L_{0,-1;4,2}-L_{4,-1;4,2})+224(L_{0,0;4,2}-L_{4,0;4,2})-3099m^2L_{0,-2;4,2}=0.
\end{equation}

\textbf{Remark 3.} As in the case of the quadrupolar obstruction
(\ref{obstruction_b442}), the octupolar obstruction
(\ref{obstruction_b563}) is purely due to the non-conformal flatness
of the set up and did not arise in the analysis given in \cite{Val04d}
as the initial data considered in that case was constructed so as to
satisfy it automatically.

\subsection{Solutions at order $p=6$}

The expansions at order $p=6$ are both quantitatively and qualitatively
more complex than those of the previous orders. Firstly, we discuss 
which conditions are needed to ensure the existence of
solutions. These are given in the next

\begin{lemma}
Consider initial data satisfying the assumptions 0-5, and such that
that the conditions (\ref{violation_b342}), (\ref{violation_b463}) and
(\ref{violation_b584}) hold. Furthermore, assume that the initial data
is such that the obstructions (\ref{obstruction_b442}),
(\ref{obstruction_b542}) and (\ref{obstruction_b563}) vanish. Then the
solutions to the $p=6$ v transport equations are polynomial in $\tau$
and the $p=6$ transport equations implied by the Bianchi constraint
equations are satisfied if and only if
\begin{equation} \label{violation_b6105}
L_{0,1;10,5}=-L_{4,1;10,5}.
\end{equation}

\end{lemma}

\bigskip
The structure of the solutions to the $p=6$ Bianchi transport equations is described in the following theorem.

\begin{theorem}
Consider initial data sets satisfying the assumptions 0-5, and such 
that the conditions (\ref{violation_b342}), (\ref{violation_b463}),
(\ref{violation_b584}) and (\ref{violation_b6105})
hold. Furthermore, assume that the initial data is such that the
obstructions (\ref{obstruction_b442}), (\ref{obstruction_b542}) and
(\ref{obstruction_b563}) vanish. Then the coefficients
$a_{j,6;2q,q}(\tau)$, with $j=0,\ldots,4$, $q=|2-j|,\ldots,6$, $q\neq
2$, $q \neq 3$ have polynomial dependence in $\tau$. On the other hand,
\begin{equation}
a_{j,6;4,2}=\Upsilon^+_{6;4;k}(1-\tau)^{8-j}\mathcal{P}_j(\tau)\ln(1-\tau)+ \Upsilon^-_{6;4,2}(1+\tau)^{4+j}\mathcal{P}_{4-j}(\tau)\ln(1+\tau)+\mathcal{Q}(\tau),
\end{equation}
where the obstructions are given by 
\begin{subequations}
\begin{eqnarray}
&&\Upsilon^+_{6;4,2}=\frac{7722}{7}\mbox{i}mJQ_3-\frac{2198208}{6965}m^2J^2-\frac{20817}{14}AmJ^2 \nonumber \\
&& \phantom{XXXXX}+\frac{62691}{4816}\sqrt{6}Am\bigg( L_{4,-1;4,2}-L_{0,-1;4,2} \bigg) 
-\sqrt{6}m^2\bigg( \frac{7559711}{126420}L_{4,-1;4,2}+\frac{8613019}{126420}L_{0,-1;4,2}\bigg)  \nonumber \\
&& \phantom{XXXXX}+\frac{58}{43}\sqrt{6}A\bigg(L_{4,0;4,2}-L_{0,0;4,2}\bigg)+\sqrt{6}m\bigg( \frac{3282401}{27090}L_{4,0;4,2}+\frac{3648769}{27090}L_{0,0;4,2}\bigg) \nonumber \\
&& \phantom{XXXXX}+\frac{144}{13}\sqrt{6}\bigg(L_{4,1;4,2}+L_{0,1;4,2}\bigg) \nonumber\\
&& \phantom{XXXXX} -\frac{11500}{91}\sqrt{6}AmY^+_{3;4,2}-\frac{2985}{56}\sqrt{6}mY^+_{4;4,2}+\frac{18}{35}\sqrt{6}(Z^-_{5;4,2}-Y^+_{5;4,2}), \label{obstruction_b642a}\\
&& \Upsilon^-_{6;4,2}=\frac{7722}{7}\mbox{i}mJQ_3+\frac{2198208}{6965}m^2J^2-\frac{20817}{14}AmJ^2  \nonumber \\
&& \phantom{XXXXX}+\frac{62691}{4816}\sqrt{6}Am\bigg( L_{4,-1;4,2}-L_{0,-1;4,2} \bigg) 
-\sqrt{6}m^2\bigg( \frac{7559711}{126420}L_{0,-1;4,2}+\frac{8613019}{126420}L_{4,-1;4,2}\bigg) \nonumber \\
&& \phantom{XXXXX}+\frac{58}{43}\sqrt{6}A\bigg(L_{4,0;4,2}-L_{0,0;4,2}\bigg)+\sqrt{6}m\bigg( \frac{3282401}{27090}L_{0,0;4,2}+\frac{3648769}{27090}L_{4,0;4,2}\bigg) \nonumber\\
&& \phantom{XXXXX}+\frac{144}{13}\sqrt{6}\bigg(L_{4,1;4,2}+L_{0,1;4,2}\bigg) \nonumber \\
&& \phantom{XXXXX} -\frac{11500}{91}\sqrt{6}AmY^+_{3;4,2}-\frac{2985}{56}\sqrt{6}mY^+_{4;4,2}+\frac{18}{35}\sqrt{6}(Z^-_{5;4,2}-Y^+_{5;4,2}). \label{obstruction_b642b}
\end{eqnarray}
\end{subequations}
Moreover,
\begin{equation}
a_{j,6;6,3}(\tau)=\Upsilon_{6;6,3}\big( (1-\tau)^{8-j}\mathcal{P}_{j+1}(\tau)\ln(1-\tau)+(1+\tau)^{4+j}\mathcal{P}_{5-j}(\tau)\ln(1+\tau) \big) +\mathcal{Q}(\tau),
\end{equation}
where
\begin{eqnarray}
&&\hspace{-15mm}\Upsilon_{6;6,3}=24m^3w_{3,6,3}-\frac{565753248}{82585}\mbox{i}m^2J^3 \nonumber \\
&&\hspace{-3mm}+\frac{36399}{830}\sqrt{6}\mbox{i}m^2 J\bigg(L_{4,-1,4,2}-L_{0,-1;4,2}\bigg)+\frac{1904}{415}\sqrt{6}\mbox{i}m J\bigg(L_{4,0;4,2}-L_{0,0;4,2}\bigg) \nonumber \\
&&\hspace{-3mm} -\frac{7272}{415}\sqrt{30}mL_{0,-1,6,3}+\frac{408}{83}\sqrt{30}m^2\bigg(L_{0,0;6,3}-L_{4,0;6,3}\bigg)+\frac{576}{415}\sqrt{30}\bigg( L_{0,1;6,3}-L_{4,1;6,3}\bigg) \nonumber\\
&&\hspace{-3mm}-\frac{233216}{415}\sqrt{6}\mbox{i}m JY^+_{3;4,2}+6\sqrt{30}Y^+_{4;6,3}. \label{obstruction_b663}
\end{eqnarray}

\end{theorem}

\textbf{Remark 1.} The most noticeable feature of the order $p=6$
quadrupolar obstructions (\ref{obstruction_b642a}) and
(\ref{obstruction_b642b}) is that they are time asymmetric. That is,
the vanishing of one of them  does not imply the vanishing of the other at
least if $J\neq 0$ ---this feature had already appeared in the
conformally flat case discussed in paper I. However, it is noted that
at least to this order the conformal flatness plays no role in the
time asymmetric behaviour.

\textbf{Remark 2.} If both $\Upsilon^+_{6;4,2}$ and
$\Upsilon^-_{6;4,2}$ vanish then the expressions
(\ref{obstruction_b642a}) and (\ref{obstruction_b642b}) can be used to
solve for, say, $L_{0,-1;4,2}$ and $L_{4,-1;4,2}$ in a similar way to
how it was done in paper I.

\section{Conjecture about the general structure of the solutions}

The results of the CA calculations carried out in the previous section
exhibit a well defined structure. From here and together with the
results of part I, it is not hard to conjecture what the behaviour of
the solutions for a generic order $p$ should be.

Firstly, we note that the conditions (\ref{violation_b342}),
(\ref{violation_b463}), (\ref{violation_b584}) and
(\ref{violation_b6105}) arise, arguably, as a side effect of the way
we have calculated expansions consistent with the momentum constraint
(\ref{momentum}) and that strictly speaking we have not found
solutions. It is quite likely that the use of explicit solutions to
the momentum constraint ---as it is possible in the axially symmetric
case, see section \ref{section:existence}--- would get rid of such
conditions. Therefore, conditions of this sort will be disregarded in
the subsequent discussion. Moreover, following the general trend of
the paper, we limit ourselves to the axially symmetric case. The
generalisation of the discussion given in the sequel for non-axially
symmetric initial data sets should be direct.

As with the expansions carried out in section 5, we restrict ourselves
to a class of initial data sets satisfying assumptions 0-5.  Now,
assuming that for a given integer $p_*\geq 7$ all lower
order obstructions vanish  ---i.e. those arising from the $p=4,\ldots,p_*-1$
Bianchi constraint equations--- it is conjectured that the solutions
to the Bianchi transport equations are such that:

\begin{itemize}

\item[(i)] The coefficients
\begin{equation}
a_{j,p;2p,p}(\tau), \quad a_{j,p;2(p-1),p-1}(\tau), \quad j=0,\ldots,4,
\end{equation}
associated with the harmonics $\TT{2p}{p}{p+j}$ and $\TT{2(p-1)}{p-1}{p-1+j}$, respectively, are polynomials in $\tau$.

\item[(ii)] The coefficients $a_{j,p;2(p-2),p-2}$ accompanying the harmonics $\TT{2(p-2)}{p-2}{p-2+j}$ are of the form
\begin{eqnarray}
&&\hspace{-3cm}a_{j,p;2(p-2),p-2}(\tau)=\Upsilon_{p;2(p-2),p-2}\bigg( (1-\tau)^{p+2-j}\ln(1-\tau)\mathcal{P}_{p-4+j}(\tau) \nonumber \\
&&\hspace{4cm}+(1+\tau)^{p-2-j}\ln(1+\tau)\mathcal{P}_{p-j}(\tau)\bigg) +\mathcal{Q}(\tau),
\end{eqnarray}
where the obstruction $\Upsilon_{p;2(p-2),p-2}$ is given by
\begin{equation}
\Upsilon_{p;2(p-2),p-2}=Z^-_{p-1;2(p-2),p-2}-Y^+_{p-1;2(p-2),p-2}.
\end{equation}

\item[(iii)] The coefficients $a_{j,p;2(p-3),p-3}$ which go together
with the harmonics $\TT{2(p-3)}{p-3}{p-3+j}$ are of the form
\begin{eqnarray}
&&\hspace{-3cm} a_{j,p;2(p-3),p-3}(\tau)=\Upsilon_{p;2(p-3),p-3}\bigg( (1-\tau)^{p+2-j}\ln(1-\tau)\mathcal{P}_{p-5+j}(\tau) \nonumber \\
&&\hspace{4cm} +(1+\tau)^{p-2+j}\ln(1+\tau)\mathcal{P}_{p-1-j}(\tau)\bigg) +\mathcal{Q}(\tau),  
\end{eqnarray}
where $\Upsilon_{p;2(p-3),p-3}$ is a time symmetric obstruction
similar in structure to those given by (\ref{obstruction_b542}) and
(\ref{obstruction_b663}), and should be a quantity of $2^{p-3}$-polar
nature. That is, it should be possible to write it as a linear
combination of quantities quadrupolar nature like
\[
w_{p-3;2(p-3),p-3}, \quad Y^+_{p-2;2(p-3),p-3} , \quad (L_{4,s;2(p-3),p-3}-L_{0,s;2(p-3),p-3}),
\]
and $2^{p-3}$-polar products of lower multipole quantities such as
\[
J^{p-3}, \quad JY^+_{p-3;2(p-4),p-4}, \quad J(L_{4,s;2(p-4),p-4}-L_{0,s;2(p-4),p-4}).
\]

\item[(iv)] The coefficients $a_{j,p,2q,q}$ for $q=2,\ldots,p-4$ have
also logarithmic dependence, being of the form
\begin{eqnarray}
&& \hspace{-3cm}a_{j,p;2q,q}(\tau)=\Upsilon^+_{p;2q,q}(1-\tau)^{p+2-j}\ln(1-\tau)\mathcal{P}_{q+j-2}(\tau) \nonumber \\
&&\hspace{4cm} + \Upsilon^-_{p;2q,q}(1+\tau)^{p-2+j}\mathcal{P}_{q+2-j}(\tau)
\end{eqnarray}
where $\Upsilon^+_{p;2q,q}$ and $\Upsilon^-_{p;2q,q}$ are obstructions of $2^q$-polar nature. Furthermore,
\[
\Upsilon_{p;2q,q}^+=0 \nLeftrightarrow \Upsilon^-_{p;2q,q}=0,
\]
that is, the obstructions are time asymmetric.

\item[(v)] The coefficients associated with  harmonics of the form
$\TT{2}{1}{1}$ and $\TT{0}{0}{0}$, that is $a_{j,p;2,1}(\tau)$,
$j=1,2,3$ and $a_{2,p;0,0}(\tau)$ are polynomials in $\tau$.

\end{itemize}

\bigskip
Providing a proof of the above structure would constitute a
remarkable feat. As things stand now, it would require a much deeper
understanding of the group theoretical structures underlying the
cylinder at spatial infinity, $\mathcal{I}$.

\section{Conclusions}
The discussion presented in the current article is the natural
extension of the analyses carried out in
\cite{Val04a,Val04d,Val04e}. The objective at the start of this
programme was to provide evidence for a conjecture presented by
H. Friedrich in \cite{Fri03a}. As discussed at length elsewhere, this
condition stated that only mild asymptotic conditions on the initial
data were required in order to guarantee the existence of developments
with a smooth null infinity. The conjectured asymptotic requirements
were essentially that condition (i) in \ref{theorem:regularity} holds
to all orders. As shown in \cite{Val04a} this conjecture resulted too
optimistic. The results in \cite{Val04a,Val04d,Val04e} have suggested
that Penrose's notion of \emph{asymptotic simplicity} imposes stringent
restrictions on the behaviour of the gravitational field in the
so-called region of spacetime ``close to null and spatial
infinity''. Namely, that the field should be static/stationary in this
region ---a startling rigidity result if proved to be true.

The basis of this new conjecture was the discovery of a hierarchy of
so-called \emph{obstructions} to the smoothness of null
infinity. Roughly, these obstructions are quantities which can be fully
determined in terms of initial data sets and whose vanishing eliminates
some particular singular logarithmic behaviour at the sets where null
infinity ``touches'' spatial infinity. Some of these obstructions can
be fully written in terms of the freely specifiable data ---like the
ones appearing in section (ii) and (iv) of section 9--- while others
involve the coefficients $w_{p;2q,q}$ which arise from the expansions
of the conformal factor of the initial hypersurface $\vartheta$. The
latter class of obstructions is much more complex, as it is not clear
---at least at first sight--- how can one construct data for which
these obstructions vanish; the reason being that the coefficients
$w_{p;2q,q}$ are not freely specifiable. They are part of the solution
to an elliptic equation (at least if one uses the conformal method to
solve the constraints), and contain information of global nature. They
are, in principle, (complicated) functions of the free data.
 
In paper I, it was shown that the assumption regarding obstructions up
to a certain order ($p=8$) vanish leads naturally to initial data sets
which are asymptotically Schwarzschildean. Ideally, in the present
paper, one would like to repeat the same type of procedure with the
obstructions we have been calculated, however they seem to be
insufficient to be able to extract some type of statement concerning
asymptotic stationarity. The way stationary solutions fit in the whole
scheme, will be the subject of further study.

Finally, we would like to make a remark concerning a point that has
been implicit through the whole programme carried out in
\cite{Val04a,Val04d,Val04e} and being concluded in the present
article. Namely, that the hierarchy of obstructions to the smoothness
we have been discussed are also \emph{obstructions to the peeling
behaviour of the spacetimes}, at least in what it concerns the lower
order obstructions: $\Upsilon_{4;4;2}$, $\Upsilon_{5;4,2}$,
$\Upsilon_{5;6,3}$. Moreover, there are reasons to believe that the
constancy of the Newman-Penrose constants would involve the
obstructions arising at order $p=6$ ---i.e. $\Upsilon^\pm_{6;4,2}$,
$\Upsilon_{6;6,3}$ and $\Upsilon_{6;8,4}$. The first steps to
disentangle this connection have been sketched in \cite{Val04b}. A
more complete discussion of this connection will be given elsewhere.

\section*{Acknowledgments}
I want to thank R. Beig, H. Friedrich for encouragement and helpful
discussions. Further gratitude is due to S. Dain for insightful
conversation and for sharing unpublished results which give rise to
part of the content of section 2 ---theorem \ref{existence_psi} and
\ref{existence_theta}. Gratitude is also due to CM Losert for a careful
reading of the manuscript. The computer algebra which constitute the
core of the investigation were carried out in the computers of the Max
Planck Institute f\"ur Gravitationsphysik, Albert Einstein Institute
in Golm, Germany. This research is funded by the Austrian FWF (Fonds
zur F\"{o}rderung der Wissenschaftliche Forschung) via a Lise Meitner
fellowship (M690-N02 \& M814-N02).


\end{document}